\documentstyle[11pt,aaspp4]{article}

\begin{document}

\begin{center}
{\it Astrophysical Journal Letters}, 1996, in press
\end{center}

\title{Physical Conditions in Low Ionization Regions of the 
Orion Nebula\altaffilmark{1}}

\author{J.~A.~Baldwin\altaffilmark{2}, 
A.~Crotts\altaffilmark{3}, 
R.~J.~Dufour\altaffilmark{4}, 
G.~J.~Ferland\altaffilmark{5}, 
S.~Heathcote\altaffilmark{2}, 
J.~J.~Hester\altaffilmark{6}, 
K.~T.~Korista\altaffilmark{5}, 
P.~G.~Martin\altaffilmark{7},  
C.~R.~O'Dell\altaffilmark{4}, 
R.~H.~Rubin\altaffilmark{8,9}, 
A.~G.~G.~M.~Tielens\altaffilmark{8}, 
D.~A.~Verner\altaffilmark{5}, 
E.~M.~Verner\altaffilmark{5}, 
D.~K.~Walter\altaffilmark{10}, and 
Z.~Wen\altaffilmark{4}}

\altaffiltext{1}{Based in part
on observations made with the NASA/ESA {\it Hubble Space
Telescope}, obtained at the Space Telescope Science Institute, which is 
operated by AURA, Inc., under NASA contract NAS5-26555}  
\altaffiltext{2}{Cerro Tololo Inter-American Observatory, Casilla 603,
La Serena, Chile}
\altaffiltext{3}{Columbia University, Department of Astronomy, 538 W. 120th
Street, New York, NY 10027}
\altaffiltext{4}{Rice University, Space Physics \& Astronomy, Box 1892,
Houston, TX 77251-1892}
\altaffiltext{5}{University of Kentucky, Physics \& Astronomy, Lexington, 
KY 40503-0055}
\altaffiltext{6}{Arizona State University, Physics \& Astronomy, Tempe,
AZ 85287-1504}
\altaffiltext{7}{Canadian Institute for Theoretical Astrophysics, University 
of Toronto, Toronto, ON, Canada M5S~3H8}
\altaffiltext{8}{NASA/Ames Research Center, Moffett Field, CA 94035}
\altaffiltext{9}{Orion Enterprises}
\altaffiltext{10}{South Carolina State University, Department of Physical
Sciences, Box 7296, Orangeburg, SC 29117}

\begin{abstract}
We reexamine the spectroscopic underpinnings of recent suggestions that 
[\ion{O}{1}] and [\ion{Fe}{2}] lines from the Orion
\ion{H}{2} region are produced in gas where the iron-carrying grains
have been destroyed and the electron density is surprisingly high. Our
new observations show that previous
detections of [\ion{O}{1}] 5577 were dominated by telluric emission.
Our limits are consistent with a moderate density ($\approx
10^4$~cm$^{-3}$) photoionized gas. We show that a previously proposed
model of the Orion \ion{H}{2} region reproduces the observed
[\ion{O}{1}] and [\ion{Fe}{2}] spectrum.  These lines are fully
consistent with formation in a dusty region of moderate density.
\end{abstract}


\section{Introduction}

The Orion Nebula is the defining blister \ion{H}{2} region
(\cite{zuc73}; \cite{bal74}). A star cluster ionizes the skin of the
molecular cloud, causing an expansion away from the molecular cloud
towards us.  The \ion{H}{2} region is in
photoionization equilibrium with a
density in concert with that of the background
photodissociation region (\cite{tie85}), and both regions are dusty with
high depletions of the refractory elements (\cite{rub93}). 

Bautista, Pradhan, \& Osterbrock (1994;
hereafter BPO), Bautista \& Pradhan (1995; hereafter BP95), and
Bautista, Peng, \& Pradhan (1996; hereafter BPP) have recently 
suggested that
[\ion{O}{1}] and [\ion{Fe}{2}] lines are produced by a warm ($T_e
\approx 10^4$~K) region with very high electron density ($n_e \sim 2
\times 10^6$~cm$^{-3}$) and a solar iron abundance.  This would be surprising
for such a dusty
environment, but would have important implications for grain destruction
mechanisms.
Here we use better spectroscopic data and new photoionization
calculations to reexamine these claims.  

\section{The [O I] Spectrum}

BP95's claims 
were based in part on their analysis of the [\ion{O}{1}]
intensities reported by Osterbrock, Tran, \& Veilleux (1992; hereafter
OTV) and Baldwin et al.\ (1991; hereafter BFM). 
In these low-resolution spectra the 
Orion [\ion{O}{1}] lines were completely blended with the 
telluric emission which varies with time and position on the sky. 
In the case of [\ion{O}{1}] 5577, the telluric emission is much stronger
than nebular emission, and
no careful subtraction could yield an accurate measurement of this line.
Both BFM and OTV 
assigned large uncertainties to their measurement of [\ion{O}{1}] 6300, and 
did not claim any significant measurement of [\ion{O}{1}] 5577.

We have two sets of observations which dramatically improve on these 
measurements. First, we used the Faint Object
Spectrograph on the {\it Hubble Space Telescope (HST)} with the
0.86{\arcsec} circular aperture and the G570H grating
on 1995 October 23/24 (UT).
We made 225~s integrations  
at positions 1SW ($\alpha_{2000}$, $\delta_{2000}$) =
($05^{\rm h}35^{\rm m}14\fs71$, --05$^{\rm o}$23\arcmin41\farcs5) and
x2 ($05^{\rm h}35^{\rm m}16\fs92$, --05$^{\rm o}$23\arcmin57\farcs5),
roughly 30{\arcsec} WSW and S, respectively, from $\theta^1$ Ori C. The
summed spectrum is presented in Figure 1. Line strengths were
measured with Gaussian fits to the profiles and are presented in Table
1 where they are given relative to \ion{He}{1} 6678 and are uncorrected
for interstellar extinction.  We note that the {\it HST} spectra
are not affected by telluric emission. The diode glitch between 
[\ion{Cl}{3}] 5538 and
[\ion{O}{1}] 5577 in the flat fields does not impact our
analysis. 

Also, a spectrum was taken with the Cassegrain echelle spectrograph on
the 4m Blanco Telescope
at CTIO,
on 12 January 1996. This provided an
11~km~s$^{-1}$ resolution spectrum at a time when the Earth's motion
split the nebular lines from the telluric emission.  The spectra were
extracted over a slit width of 1{\arcsec} and length of 12.5{\arcsec},
centered on the BFM position 1 at 37{\arcsec} W of $\theta^1$ Ori C.
The line intensities are in Table 1 and parts of the spectra are shown
in Figure 1. 

Orion has a spectrum with a strong point to point
variation (e.g., \cite{pei77}) causing the
differences in the two spectra.  
The CTIO spectrum at the
BFM position has the
higher ionization. The $HST$ spectrum is
similar to that reported by OTV for another position in the
nebula.  
Table 2 gives reddening
corrected intensities relative to H$\beta$, using BFM's extinction
for the CTIO data; for our {\it HST} data, we adopt
C(H$\beta)=0.51$ from the composite H$\alpha$/H$\beta$ ratio. The CTIO
data in Table 2 were renormalized using the \ion{He}{1} 6678/H$\beta$
intensity ratio reported by BFM.

Limits to the [\ion{O}{1}] ratio $R \equiv I(6300+6364)/I(5577)$ are
needed to set $n_e$ for an assumed $T_e$. In neither spectrum 
was [\ion{O}{1}] 5577 detected, but we have obtained a much
stricter upper limit.  
The observed limits are
$R>84$ (CTIO) and $R>33$ ({\it HST}) and the reddening corrected
ratios are $R>71$ and $R>29$. The superior spectral
resolution is the primary reason the CTIO spectrum has the more
stringent limit. This limit is very different from the ratio of
22 adopted by BP95, and shows that warm dense gas does not
contribute to the [\ion{O}{1}] spectrum.

The CTIO [\ion{S}{2}] 6716/6731 ratio (0.53)
indicates a $T_e$-insensitive density of $\sim 6600$~cm$^{-3}$
(\cite{ost89}). The [\ion{N}{2}] 5755/6583 ratio (0.018) indicates
$T_e$ of 10,100~K, fairly independent of $n_e$ in this range.  The {\it
HST}-derived properties are similar.  Assuming $10^4$~K, the CTIO
[\ion{O}{1}] ratio limit sets the limit $n_e<1.6 \times
10^5$~cm$^{-3}$, consistent with the $n_e$[\ion{S}{2}] and much lower
than the density deduced by BP95.  Assuming $n_e$[\ion{S}{2}], we
find that the [\ion{O}{1}] ratio limit sets the limit $T_e < 11,600$~K,
consistent with the [\ion{N}{2}] spectrum.  These are all conventional
Orion \ion{H}{2} region numbers; we conclude that the [\ion{O}{1}]
spectrum is consistent with its formation in gas of moderate density
($\sim 10^4$~cm$^{-3}$). The photoionization calculations presented
below predict the [\ion{O}{1}] ratio to be $R \approx 100$.

\section{The [Fe II] Spectrum}

Table 3 shows the observed (extinction corrected) [\ion{Fe}{2}] line
ratios from OTV and Lowe et al.\ (1979) and some predictions.  
The claim by BPO that the
[\ion{Fe}{2}] lines come from a high-density region in Orion
is based on a collisional model of the \ion{Fe}{2} atom, which
reproduces the observed optical to
near-infrared line ratios at density $n_e \sim 10^6$~cm$^{-3}$.  Table 3
shows the collisional model reproduced from Table 2 of BPP
($n_e=2 \times 10^6$~cm$^{-3}$ and $T_e=10^4$~K, using new \ion{Fe}{2}
collision strengths described by Bautista \& Pradhan 1996, hereafter BP96).
The predictions of the collisional model are sensitive to the adopted
collision strengths. 
Our model~I recalculates the line ratios for the same collisional model,
using the publicly available collisional data from Zhang \& Pradhan
(1995; hereafter ZP) also used in our photoionization models below. This
shows that the lines 4815, 5159, 5262, and 5334 are stronger by a factor
of two than in the BP96-based model in BPP because of different collision
strengths.
Also shown in Table 3 are line ratios we calculated for a density more
typical of standard Orion models, $n_e=10^4$~cm$^{-3}$ (model II).
At this density, there is clearly a requirement for some mechanism other
than simply collisions to populate the upper levels of the \ion{Fe}{2}
that produce the optical lines.

The alternative to collisional excitation, pumping of [\ion{Fe}{2}]
lines by the incident continuum, was considered by BPP (see also Lucy
1995).  The equation used by BPP is correct in the optically thin limit.
However, the ultraviolet lines dominating the pumping are usually
optically thick and line self-shielding must be included if energy is to
be conserved (Ferland 1992). The correct formulation of the constant 
temperature
problem outlined in BPP should include the column density as an
additional free parameter.

To model the production of the [\ion{Fe}{2}] lines more realistically,
we incorporated a 371 level \ion{Fe}{2} model atom into Cloudy
(\cite{fer96}, \cite{ver96}).  Collision data for our model atom 
are from ZP, radiative data mainly from Quinet, Le Dourneuf, \& Zeippen
(1996) and Nahar (1995), and energy levels from Johansson (1994).  A
cloud is divided into $\sim 10^2$ zones, the ionization and $T_e$ are
determined self-consistently, and the emission is determined for the
local optical depths.  We developed the code to model the broad line
regions of quasars (Verner et al.\ 1995, 1996), incorporating the
physics described by Wills, Netzer, \& Wills (1985).  We included
pumping by line overlap from all $\approx 10^4$ emission lines included
in Cloudy, by \ion{Fe}{2} line overlap, and by the incident continuum.
We verified that the \ion{Fe}{2} atom goes to LTE in the high-density
limit, and that in the low density limit each excitation is followed
simply by radiative cascades.
A simple limit of our \ion{Fe}{2} model (one zone, constant temperature,
constant density, no continuum radiation flux, no line pumping, no
interactions with other species) corresponds to the BPO and BPP
collisional model, and we
can reproduce in this limit the \ion{Fe}{2} line ratios presented in
Figure 2 of BPO, using their atomic data.  Note that Figure 2$b$ of BPO
erroneously compared the predicted $I(8892)/I(5334)$ ratio with the
observed inverted $I(5334)/I(8892)$ ratio, although their caption for
Figure 2$b$ stated the ratio to be $I(5262)/I(5159)$.

Tables 2 and 3 present results of our photoionization calculations.
Model A is calculated with the parameters taken from BFM with improved
gas-phase elemental abundances based on results from OTV and Rubin
et al.\ (1991, 1992): H:He:C:N:O:Ne:Na:Mg:Al:Si:S:Cl:Ar:Ca:Fe
$=1:0.095:3 \times 10^{-4}:7 \times 10^{-5}:4 \times 10^{-4}:6 \times
10^{-5}:3 \times 10^{-7}:3 \times 10^{-6}:2 \times 10^{-7}:4 \times
10^{-6}: 10^{-5}:10^{-7}:3 \times 10^{-6}:2 \times 10^{-8}:3 \times
10^{-6}$, grains, ionization by the central star $\theta^1$~Ori~C, and a
hydrostatic blister.  Results of model A differ from the original BFM
model, since the underlying atomic database has been improved
considerably.  Model B has the same parameters as model A, but the
incident continua from the other three Trapezium stars are included (an
estimated 70\%
of the total flux at 2300~\AA\ is due to $\theta^1$ Ori C). In models A
and B, the gas pressure was kept constant.  Model C is the same as model
B, except that the total (gas and radiative) pressure is kept constant;
this reveals some sensitivity of the predicted spectrum to the assumed
pressure law.

Table 3 highlights the [\ion{Fe}{2}] line ratios obtained from our
photoionization models. A comparison with the observed line ratios and
with the line ratios calculated within the collisional model at the same
density $n_e=10^4$~cm$^{-3}$ (model II) indicates that the upper levels
of the \ion{Fe}{2} atom, those that produce the optical lines, can be
effectively populated by continuum pumping. The lower levels that
produce the 8617, 8892, and 12567 lines are mostly populated by
collisions in both photoionization and collisional models.  Note
therefore the sensitivity of the optical to near-infrared [\ion{Fe}{2}]
line ratios to the increased pumping from the other Trapezium stars
(model B versus model A).

Further evidence against high electron densities in the [\ion{Fe}{2}] 
emitting region in Orion comes from infrared observations
by Lowe et al. (1979). They find that the 12567 line is the only \ion{Fe}{2} 
line clearly detected in the infrared spectrum of Orion -- it is actually the
strongest [\ion{Fe}{2}] line in the infrared and optical.  Our models at
$n_e \sim 10^4$~cm$^{-3}$ predict that the 12567 line is $\sim 3$ times {\it
stronger} than the 8617 line, in good agreement with the
observations. If the 12567 line were 2--3 times {\it weaker} than the
8617 line as predicted at $n_e \sim
10^6$~cm$^{-3}$, it would not have been detected at all.

Table 2 includes the relative strength of [\ion{Fe}{2}] 8617 compared to
lines of other species, information crucial to abundance analysis.  Our
photoionization models match the observations in Tables 2 \& 3
quite well.  The strongly
depleted Fe/H in the gas phase, $3 \times 10^{-6}$, is close to the
value $\sim 2.7 \times 10^{-6}$ found from KAO observations of the
ground state [\ion{Fe}{3}] 22.9$\mu$m line (\cite{eri89} and in
preparation).  Even with solar abundances, iron is not a major coolant
in Orion, and so the intensities of [\ion{Fe}{2}] lines scale linearly
with abundance.  Therefore, if solar Fe/H (BPO, BP95) were used instead, the
predicted [\ion{Fe}{2}] lines would be 11 times stronger and very
discrepant with observations.

BP95, using [\ion{Fe}{2}]/[\ion{O}{1}] line ratios, inferred that 
the Fe/O abundance in the proposed high-density
region is close to solar.
Based on our new upper limit on the [\ion{O}{1}] 5577 strength, 
which is 3.2 times lower than one used by BP95,
the Fe/O abundance
would be higher than 3 times solar, if we adopt their high density.
If iron were depleted, 
these numbers would be even larger. 

Both photoionization and
collisional models predict that the
[\ion{Fe}{2}] and [\ion{O}{1}] emission line regions should be nearly
coincident. Our upper limit on the density derived from the [\ion{O}{1}]
line ratio is consistent with the origin of [\ion{Fe}{2}] lines in
a region of
moderate density, $n_e \sim 10^4$~cm$^{-3}$.

\section{Summary}

1. Previous measurements of [\ion{O}{1}] 5577 were affected by telluric
emission.  Our limit to the line sets limits on the density and
temperature consistent with the physical conditions inferred from the
other emission lines.

2. The high-density collisional model for the formation of the
[\ion{Fe}{2}] lines is inconsistent with the strongest observed
[\ion{Fe}{2}] line at 12567~\AA.  
In combination with the new upper limit on the [\ion{O}{1}] 5577 line, the
inferred Fe/O ratio in any putative high-density region
would be at least 3 times solar, not solar as
found by BP95.

3. Our photoionization models of the [\ion{Fe}{2}] spectrum with
parameters close to those found by BFM predict a spectrum in good
overall agreement with observations. The infrared [\ion{Fe}{2}] lines
are produced by collisional excitation, and the optical [\ion{Fe}{2}]
lines by photon pumping by the ultraviolet continuum in a region of
moderate density, $n_e \sim 10^4$~cm$^{-3}$. Fe/H must be depleted,
$\sim 3 \times 10^{-6}$, in agreement with independent 
determinations from KAO observations.

4. While dense condensations undoubtably exist within the Orion Nebula,
these regions do not contribute significantly to the [\ion{Fe}{2}] and
[\ion{O}{1}] spectra.

The photoionization models have several parameters which can affect the
resulting [\ion{Fe}{2}] spectrum, including the flux of the pumping
continuum, turbulent velocity, and bulk motions in the gas.  The shape
of the ionizing continuum, the pressure law, and gas phase abundances
are major parameters which affect the emission of the stronger CNO
lines.  Detailed fitting of the Orion Nebula spectra will be the subject
of a future paper.

\acknowledgments
We thank STScI for its support through
grants GO-4385, GO-5748,  GO-6056, GO-6093, and AR-06403.01-95A.  RJD
and GJF acknowledge NASA/Ames Research Center Interchange Grants
NCC2-5008 and NCC2-5028, respectively.  Research in Nebular
Astrophysics at the University of Kentucky is supported by the National
Science Foundation through award AST 93-19034.  We are grateful to 
D. Osterbrock, L. Lucy, A. Pradhan and M. Bautista for useful discussions.

\clearpage
 
\begin{deluxetable}{llll}
\footnotesize
\tablecaption{New Observations}
\tablewidth{0pt}
\tablehead{
\colhead{Ion} & \colhead{Wavelength} & \colhead{CTIO\tablenotemark{a}} 
& \colhead{$HST$\tablenotemark{b}}
} 
\startdata
[Cl III] & 5518 & -- & 0.106 \nl
[Cl III] & 5538 & 0.13 & 0.119 \nl
[O I] & 5577 & $<0.003$ & $<0.0128$ \nl
[O I] & 5577 sky & 0.044 & -- \nl
[NII] & 5755 & 0.17 & 0.322 \nl
He I & 5876 & 3.22 & 3.44 \nl
[O I] & 6300 & 0.19 & 0.313 \nl
[O I] & 6300 sky & 0.0071 & -- \nl
[S III] & 6312 & 0.49 & 0.565 \nl
Si II & 6347 & 0.042  & -- \nl
[O I] & 6364 & 0.063 & 0.113 \nl
Si II & 6372 & 0.022 & -- \nl
[N II] & 6548 & 3.30 & 7.35 \nl
H$\alpha$ & 6563 & 76-113\tablenotemark{c} & 85.2 \nl
[N II] & 6583  & 11.4 & 21.0 \nl
He I & 6678 & 1.00 & 1.00 \nl
[S II] & 6716 & 0.52 & 0.865 \nl
[S II] & 6731 & 0.98 & 1.75 \nl
\enddata
\tablenotetext{a}{New CTIO echelle data at BFM position 1.  Not dereddened.}
\tablenotetext{b}{{\it HST} observations. Not dereddened.}
\tablenotetext{c}{Poorly calibrated because of broad H$\alpha$ absorption in 
standard star.}
\end{deluxetable}

\clearpage
\begin{deluxetable}{lcccccc}
\footnotesize
\tablecaption{Dereddened and Predicted Spectrum\tablenotemark{a}}
\tablewidth{0pt}
\tablehead{
\colhead{} & \colhead{} & \colhead{} & \colhead{}
& \multicolumn{3}{c}{Photoionization models\tablenotemark{c}}\\
\cline{5-7}\\
\colhead{} & \colhead{Line} & \colhead{Ref\tablenotemark{b}} & 
\colhead{Observations} 
& \colhead{A} & \colhead{B} & \colhead{C}
} 
\startdata
 HeI   & 5876      & 1 &  13.7     & 13.3   & 13.3   & 12.9   \nl
[NII]  & 5755      & 1 &  0.763    & 1.06   & 1.06   & 1.97   \nl
[NII]  & 6583      & 1 &  41.6     & 53.1   & 52.8   & 74.1   \nl
[OI]   & 5577      & 1 & $<0.0136$ & 0.0044 & 0.0043 & 0.0095 \nl
[OI]   & 6300      & 1 &  0.722    & 0.341  & 0.336  & 0.699  \nl
[OII]  & 3727      & 2 &  94       & 188    & 188    & 149    \nl
[OIII] & 5007      & 2 &  343      & 465    & 460    & 379    \nl
[SII]  & 6716      & 1 &  1.85     & 1.63   & 1.62   & 1.67   \nl
[SII]  & 6731      & 1 &  3.47     & 3.19   & 3.17   & 3.57   \nl
[SIII] & 9532      & 2 &  144.5    & 151    & 151    & 150    \nl
[FeII] & 8617\tablenotemark{d} & 3 &  0.0665   & 0.0572 & 0.0584 & 0.0979 \nl
\enddata
\tablenotetext{a}{Line intensities are scaled to $0.01I($H$\beta)$.}
\tablenotetext{b}{Data references: 1 -- new CTIO, 2 -- table 7 of BFM,
3 -- OTV.}
\tablenotetext{c}{
$n_e=10^4$~cm$^{-3}$. (A) Calculations for the parameters given by 
BFM, pumping by $\theta^1$~Ori~C. 
(B) Pumping by all four
Trapezium stars, gas pressure constant. (C) Pumping by all four
Trapezium stars, total pressure constant.}
\tablenotetext{d}{Intensities of other [FeII] lines scaled to $I(8617)$
are given in Table 3.}
\end{deluxetable}

\clearpage

\begin{deluxetable}{cccccccccc}
\footnotesize
\tablecaption{Dereddened and Predicted [Fe II] Line 
Intensities\tablenotemark{a}}
\tablewidth{0pt}
\tablehead{
 \colhead{} & \colhead{} &                           \multicolumn{4}{c}
{Collisional models} & \colhead{}
& \multicolumn{3}{c}{Photoionization models\tablenotemark{e}}\\
\cline{3-6}
\cline{8-10}\\
\colhead{Line} & \colhead{Observations\tablenotemark{b}} & 
\colhead{BPP\tablenotemark{c}} & \colhead{} &
   \colhead{I\tablenotemark{d}}
 & \colhead{II\tablenotemark{d}} &
\colhead{} & \colhead{A} & \colhead{B} & \colhead{C}
} 
\startdata
4244+4245& 0.87 & 0.96\tablenotemark{f}&& 1.3~ & 0.23&& 2.1~ & 2.4~ & 1.3~ \nl
4277     & 0.64 & 0.60                 && 0.69 & 0.10&& 0.46 & 0.50 & 0.37 \nl
4815     & 0.94 & 0.44                 && 0.96 & 0.21&& 1.1~ & 1.2~ & 0.71 \nl
5158+5159& 1.3~ & 1.0\tablenotemark{f}~&& 2.7~ & 0.62&& 1.7~ & 1.9~ & 1.1~ \nl 
5262     & 0.81 & 0.71                 && 1.4~ & 0.24&& 0.28 & 0.31 & 0.26 \nl
5334     & 0.33 & 0.53                 && 0.95 & 0.11&& 0.22 & 0.24 & 0.21 \nl
8617     & 1.00 & 1.00                 && 1.00 & 1.00&& 1.00 & 1.00 & 1.00 \nl
8892     & 0.19 & 0.40                 && 0.40 & 0.34&& 0.31 & 0.31 & 0.36 \nl
12567    & 3.3~ & \nodata              && 0.39 & 3.6~&& 3.5~ & 3.4~ & 2.6~ \nl
\enddata
\tablenotetext{a}{Line intensities are scaled to $I(8617)$.}
\tablenotetext{b}{Lowe et al.\ 1979 for 12567, OTV for all
other lines.}
\tablenotetext{c}{Collisional model from BPP; 
$T_e=10^4$~K, $n_e=2 \times 10^6$~cm$^{-3}$.}
\tablenotetext{d}{Collisional model from the present work;
$T_e=10^4$~K; I -- $n_e=2 \times 10^6$~cm$^{-3}$, II -- $n_e=10^4$~cm$^{-3}$.}
\tablenotetext{e}{(A), (B) and (C) models are the same as in Table 2.}
\tablenotetext{f}{Unresolved blends at 4244 and 5159 are not included in
the BPP model.} 
\end{deluxetable}

\clearpage
\begin{figure}
\plotfiddle{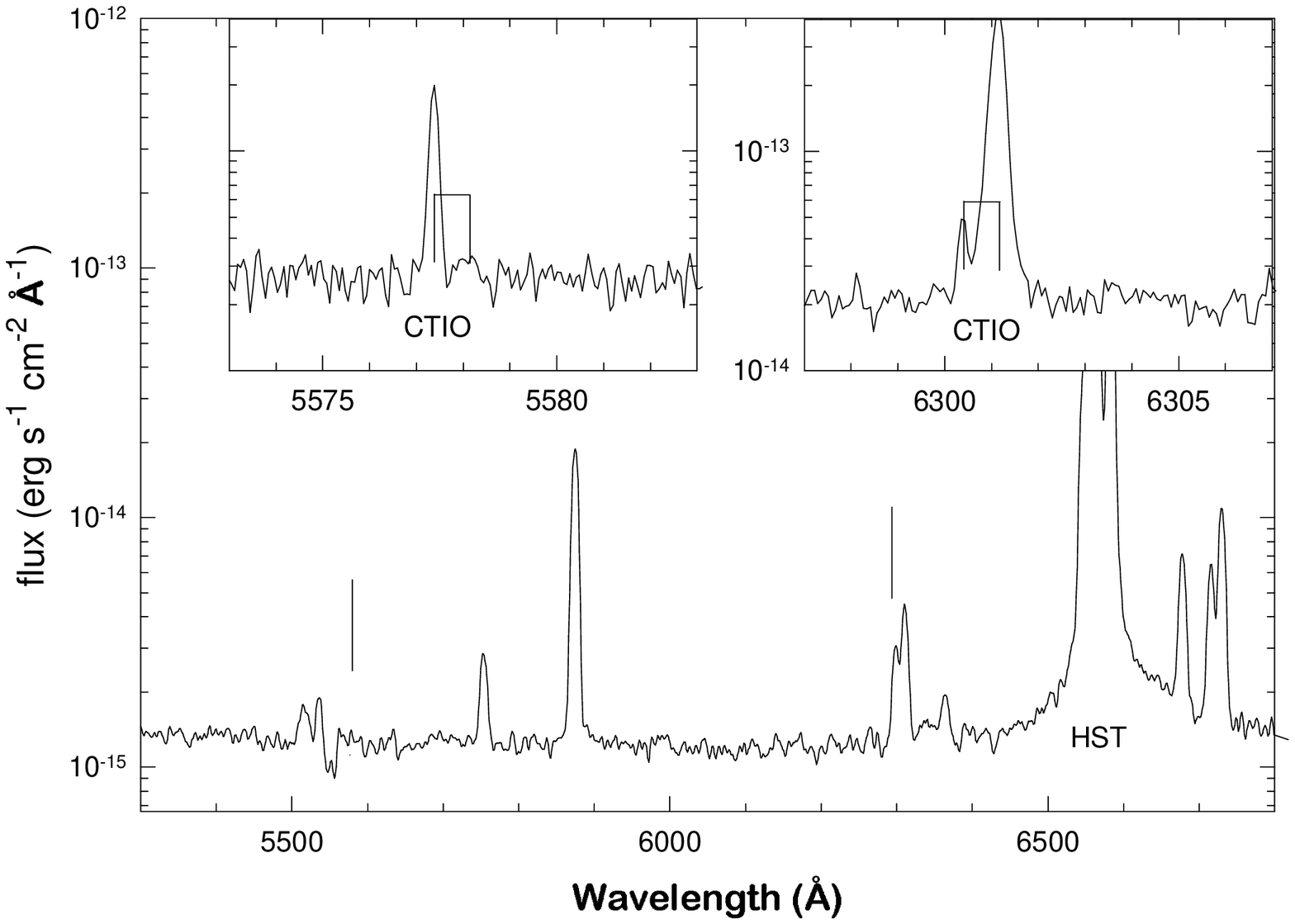}{12cm}{0}{100}{100}{-160}{250}
\caption{The spectrum of the Orion Nebula observed with the {\it HST} FOS 
and the CTIO 4m echelle. The lower half shows the {\it HST} spectrum.  The two 
inserts show portions of the echelle spectrum.  
In all spectra the [O I] lines from the Orion Nebula
are indicated by tickmarks.  In the inserts the positions of the
telluric [O I] lines in the CTIO data are also indicated, just to the
left of the lines produced in the nebula.}
\end{figure}


\clearpage

\begin{thebibliography}{}
\bibitem[Baldwin et al.\ 1991]{bal91} Baldwin, J. A., Ferland, G. J., 
     Martin, P. G., Corbin, M. R., Cota, S. A., Peterson, B. M., \& 
     Slettebak, A.\ 1991, \apj, 374, 580 (BFM)
\bibitem[Balick,  Gammon, \& Hjellming 1974]{bal74} Balick, B., 
     Gammon, R. H., \& Hjellming, R. M.\ 1974, \pasp, 86, 616
\bibitem[Bautista et al.\ 1996]{bpp96} Bautista, M. A., Peng, J., \& Pradhan,
     A. K.\ 1996, \apj, 460, 372 (BPP)
\bibitem[Bautista \& Pradhan 1995]{bau95} Bautista, M. A., \& Pradhan, 
     A. K.\ 1995, \apj, 442, L65 (BP95)
\bibitem[Bautista \& Pradhan 1996]{bau96} Bautista, M. A., \& Pradhan, 
     A. K.\ 1996, \aaps, 115, 551 (BP96)
\bibitem[Bautista et al.\ 1994]{bau94} Bautista, M. A., Pradhan, A. K., \& 
     Osterbrock, D. E.\ 1994, \apj, 432, L135 (BPO)
\bibitem[Erickson et al.\ 1989]{eri89} Erickson, E. F., Rubin, R. H., Haas, 
     M. R., Simpson, J. P., \& Colgan, S. W. J.\ 1989, BAAS, 21, 1156
\bibitem[Ferland 1992]{fer92} Ferland, G. J.\ 1992, \apj, 389, L63
\bibitem[Ferland et al.\ 1996]{fer96} Ferland, G. J., et al.\ 1996,
      \pasp, to be submitted
\bibitem[Johansson 1994]{joh94} Johansson, S.\ 1994, private communication
\bibitem[Lowe et al.\ 1979]{low79} Lowe, R. P., Moorehead, J. M., \& Wehlau, 
     W. H.\ 1979, \apj, 228, 191
\bibitem[Lucy 1995]{luc95} Lucy, L.\ 1995, \aap, 294, 555
\bibitem[Nahar 1995]{nah95} Nahar, S. N.\ 1995, \aap, 293, 967
\bibitem[Osterbrock 1989]{ost89} Osterbrock, D. E.\ 1989, Astrophysics of 
     Gaseous Nebulae and Active Galactic Nuclei (University Science Books: 
     Mill Valley, California)
\bibitem[Osterbrock et al.\ 1992]{ost92} Osterbrock, D. E., Tran, H. D., \& 
     Veilleux, S.\ 1992, \apj, 389, 305 (OTV)
\bibitem[Peimbert \& Torres-Peimbert 1977]{pei77} Peimbert, M., \& 
     Torres-Peimbert, S.\ 1977, \mnras, 179, 217
\bibitem[Quinet et al. 1996]{qui96} Quinet, P.,  Le Dourneuf, M. \& Zeippen,
     C. J.\ 1996, \aaps, in press
\bibitem[Rubin, Dufour, \& Walter 1993]{rub93} Rubin, R. H., Dufour, R. J., \& 
     Walter, D. K.\ 1993, \apj, 413, 242
\bibitem[Rubin et al.\ 1992]{rub92} Rubin, R. H., Erickson, E. F., 
     Haas, M. R., Colgan, S. W. J., Simpson, J. P., \& Dufour, R. J.\ 1992, 
     in The Astrochemistry of Cosmic Phenomena -- IAU Symposium No. 150, ed. 
     P. D. Singh (Kluwer), p. 281
\bibitem[Rubin et al.\ 1991]{rub91} Rubin, R. H., Simpson, J. P., Haas, M. R.,
     \& Erickson, E. F.\ 1991, \apj, 374, 564
\bibitem[Tielens \& Hollenbach 1985]{tie85} Tielens, A. G. G. M., \& 
     Hollenbach, D.\ 1985, \apj, 291, 722
\bibitem[Verner et al.\ 1995]{ver95} Verner, E. M.,. Ferland, G. J., Korista, 
     K. T., \& Verner, D. A.\ 1995, BAAS 27, 840
\bibitem[Verner et al.\ 1996]{ver96} Verner, E. M., et al.\ 1996, in preparation
\bibitem[Wills et al.\ 1985]{wil85} Wills, B. J., Netzer, H., \& Wills, D.\ 
     1985, \apj, 288, 94
\bibitem[Zhang \& Pradhan 1995]{zha95} Zhang, H. L., \& Pradhan, A. K.\ 1995, 
     \aap, 293, 953 (ZP)
\bibitem[Zuckerman 1973]{zuc73} Zuckerman, B.\ 1973, \apj, 183, 863
\end{thebibliography}
\end{document}